\documentclass[twocolumn,prl,aps,showpacs]{revtex4}

\usepackage{graphicx}
\usepackage{dcolumn}
\usepackage{bm}

\begin{document}

\preprint{APS/123-QED}

\title{Quasi-Particle Degrees of Freedom versus the Perfect Fluid as Descriptors of the Quark-Gluon Plasma}

\author{L.A. Linden Levy}
 \affiliation{University of Colorado, Boulder} 
\author{J.L. Nagle}
 \affiliation{University of Colorado, Boulder} 
 \email{Jamie.Nagle@Colorado.Edu}
\author{C. Rosen}
\affiliation{University of Colorado, Boulder}
\author{P. Steinberg}
\affiliation{Brookhaven National Laboratory}

\date{\today}
\begin{abstract}
The hot nuclear matter created at the Relativistic Heavy Ion Collider (RHIC) has been
characterized by near-perfect fluid behavior. 
We demonstrate that this stands in contradiction to the identification of QCD quasi-particles
with the thermodynamic degrees of freedom in the early (fluid) stage of heavy ion
collisions.
The empirical observation of constituent quark ``$n_q$'' scaling of elliptic flow~\cite{Adare:2006ti} 
is juxtaposed with the lack of such scaling behavior in hydrodynamic fluid calculations followed by 
Cooper-Frye freeze-out to hadrons.
A ``quasi-particle transport'' time stage after viscous effects break down the
hydrodynamic fluid stage, but prior to hadronization, is proposed to reconcile these
apparent contradictions. 
However, without a detailed understanding of the transitions between these stages,
the ``$n_q$'' scaling is not a necessary consequence of this prescription.
Also, if the duration of this stage is too short, it may not support
well defined quasi-particles.
By comparing and contrasting the coalescence of quarks into hadrons
with the similar process of producing light nuclei from nucleons, it
is shown that the observation of ``$n_{q}$'' scaling in the final state
does not necessarily imply that the constituent degrees of freedom
were the relevant ones in the initial state.

\pacs{25.75.Dw}

\end{abstract}

\maketitle

\section{Introduction}
The understanding that hadrons are not fundamental particles, but
rather composite systems comprised of asymptotically-free quarks and gluons,
led directly to the idea that hadronic matter under extreme conditions of
temperature and pressure would transform into a 
``quark-gluon plasma''~\cite{Shuryak:1980tp}.
In this state, quarks and gluons could be thought of as quasi-free 
particles moving along trajectories, 
such that the thermodynamic properties could be
approximated as that of an ideal gas with non-interacting degrees of
freedom.  More specifically, quarks and gluons would no longer be
confined in their parent hadron, but would be free to roam about
larger volumes.  This scenario gives 40 degrees of freedom for the
quark and gluon states (including two light
quarks~\footnote{Effectively 37 when one calculates the energy density,
due to the factor of 7/8 that arises from Fermi-Dirac statistics.}) as
opposed to 3 pion degrees of freedom in a hadronic system at a lower
temperature.  This enumeration is often associated with the order of
magnitude change in the entropy $s/T^{3}$ and energy density $\epsilon/T^{4}$ 
as observed near the transition temperature in lattice QCD calculations~\cite{Karsch:2007vw}.

This physical picture has generally been thought to be naive, since
the high densities and moderate momentum transfers should lead to
short mean free paths.  However, it has a certain persistence in the
heavy ion literature (e.g. the CERN press release~\cite{Heinz:2000bk}
or the discussion of such in the review~\cite{Muller:2006ee}) as it presents an
easily-described physical scenario with nominally well-defined
consequences.  Over the last few years, experimental data has
suggested a paradigm shift where the quark-gluon plasma is described
as a nearly inviscid fluid, and yet the picture of quasi-particles
carrying the thermodynamic degrees of freedom often persists.
Notably, recent experimental papers describing the scaling properties
of elliptic flow measurements state
that ``scaling suggests that partonic collectivity dominates the
transverse expansion dynamics~\cite{Afanasiev:2007tv}'' and
``[scaling indicates] a pre-hadronization state in which the flowing
medium reflects quark degrees of freedom~\cite{Abelev:2007rw}.''
The analysis presented in this paper suggest that these conclusions
are not well supported either by experimental data or theoretical
ideas.  

\section{Quasi-particles in the Quark-Gluon Plasma}

Interactions between propagating quarks and gluons in
the quark-gluon plasma generates thermal masses, and thus the
quarks and gluons can be called ``quasi-particles.''
A quasi-particle's mass is determined by the local properties of the
medium, which thus distinguishes between Lorentz frames.  By contrast
a ``real'' particle has a Lorentz-invariant rest mass.  
Such quasi-particle descriptions allow even strongly coupled systems
to show properties reminiscent of free non-interacting systems.  The
characteristic feature which indicates so-called ``good''
quasi-particle states is a decay width (which characterizes the
coupling strength and number of interaction channels) that is smaller
than its mass.

The dynamical role of quasi-particles in the hot phase of QCD is under intense
debate.  
%
%
In ~\cite{Blaizot:1999ip} an accurate match to lattice QCD
calculations of the entropy for a purely gluonic medium is obtained in
a perturbative calculation involving quasi-particles for temperatures
down to $2 T_c$.  They note that ``although the quasiparticle picture
suggested by such fits is a rather crude representation of the actual
physics of non-abelian gauge theories, it supports the idea that one
should be able to give accurate description of thermodynamics of the
QCD plasma in terms of its elementary excitations.''  Such
quasi-particle models have been extended and applied down to $1.05 T_c$
where they are able to describe deviations from the free quark and
gluon picture~\cite{Peshier:2005pp,Peshier:2004bv}.
However, they find that these effective degrees of freedom are rather
heavy and have a sizable width making it difficult to associate them
with ``strict'' quasi-particles.  Thus, it is not obvious that the
identification of the hot dense system with a quark-gluon plasma
necessarily implies a quasi-particle picture, even where perturbative
calculations are possible.

Additional information on quasi-particles near the transition
temperature are available from lattice QCD.  The suggestion that the
bulk thermodynamics of the quark-gluon plasma are governed by a large
set of colored bound states of light quarks~\cite{Shuryak:2004tx} has
largely been ruled out by examining fluctuations of quark number and
electric charge~\cite{Karsch:2005ps}.  Additionally,
baryon-strangeness correlations place severe limits on
$q\overline{q}$ bound states~\cite{Koch:2005vg,Majumder:2006nq}.  It
is notable that the consistency of $\chi_{ud}$ with a weakly
interacting plasma of quarks and anti-quarks above
$T_c$~\cite{Majumder:2006nq} does not provide compelling evidence that
these are well defined quasi-particles dominating the bulk
thermodynamics.

In many systems the thermodynamic degrees of freedom can be counted by
associating them with particles or quasi-particles and accounting for
their thermal motion and quantum numbers (for example $\frac{3}{2}kT$
and $\frac{5}{2} kT$ for non relativistic monatomic gases and
diatomic gases respectively).  A key question is whether or not this is the
case for the quark-gluon plasma.
The thermodynamic properties of some
systems, such as Landau's Fermi liquid~\cite{baym_landaubook}, are
dominated by the quasi-particle degrees of freedom, while in other
systems quasi-particles are a single medium excitation (for example
the fractional quantum hall effect~\cite{stormer}).  In the case of
hot QCD matter both may be interesting, but the former is the more
relevant when discussing the quark-hadron phase transition observed on
the lattice.

\section{Freeze-out from Hydrodynamics}

One of the great discoveries of the RHIC program has been that simple
hydrodynamic models provide a good description of RHIC data,
especially the development of radial and elliptic flow,
see~\cite{Huovinen:2006jp} and references therein.
The hydrodynamic equations are based on the energy-momentum tensor,
and viscosity is included as an expansion parameter in terms of
velocity gradients.  The viscosity term is set to zero for a
mathematically perfect fluid.  When invoking a picture of
quasi-particles, a connection can be made between the mean free path
of these objects via kinetic theory and the viscosity term in the
hydrodynamic equations.  The hydrodynamic limit of the quasi-particle
picture is reached when the typical gradients in the flow involve
length scales that are large compared to the typical mean free paths.


First-order estimates suggest that experimental data is consistent
with essentially zero viscosity~\cite{Teaney:2003kp}, but cannot rule
out a viscosity to entropy density ratio consistent with the lower bound
derived from the AdS/CFT duality ($\eta/s \geq
1/4\pi$)~\cite{Kovtun:2004de}.  Furthermore, estimates of collective
behavior based on classical transport (Boltzmann) approaches are
generally unable to describe RHIC data, whether due to the small cross
sections implied by a perturbative transport
picture~\cite{Molnar:2000jh}, or the large formation times required by
hadronic transport~\cite{Bleicher:2000sx}.

The hydrodynamic behavior of the medium must eventually break down as
the density of the medium drops below a threshold value, and particles
``freeze out'' into vacuum.  In addition, there may be particles (for
example at high transverse momentum) that are not equilibrated into
the bulk medium during the limited evolution time ($t<10-15$~fm/c)~\cite{Lisa:2005dd}.
Thus, in order to compare hydrodynamic calculations with experimental
data (i.e. measured hadron distributions after freeze-out),
hydrodynamic calculations are typically terminated at a chosen
temperature $T$.  This is usually chosen to be the Hagedorn
temperature $T_{ch} \sim 170$ MeV in single freeze-out models, and 
$T_{th} \sim 100$ MeV 
when a thermal freeze-out temperature is postulated due to
hadronic re-scattering.  In either case, this prescription defines a
``freeze-out hyper-surface'' in space-time, also referred to as a
``surface of last scattering.''  The Cooper-Frye
formalism~\cite{Cooper:1974mv} is then used to decay fluid elements to
hadrons by assuming local statistical hadronization.  Many
calculations also include the important non-equilibrium re-scattering
between hadrons after hadronization up to the point of
freeze-out~\cite{Bass:2000ib,Teaney:2001av}.

\begin{figure}
\includegraphics[width=\columnwidth]{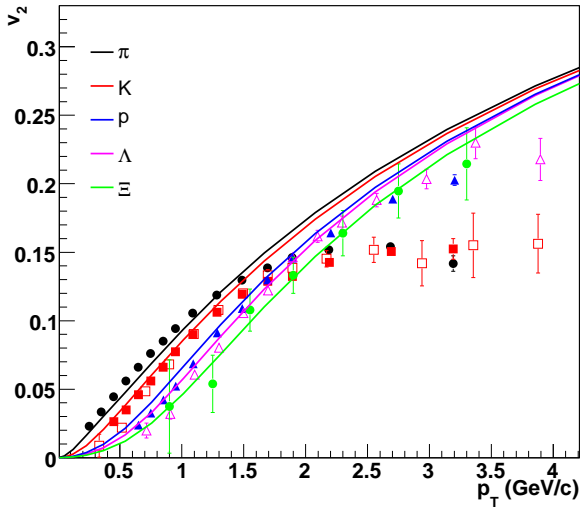}
\caption{\label{fig_v2_vs_pt_comparehydro} (color online) Elliptic
flow $v_2$ for identified hadrons as a function of $p_T$ in minimum
bias $Au+Au$ collisions at $\sqrt{s_{NN}}=200$~GeV, compared with
hydrodynamic calculations assuming Cooper-Frye freeze-out.}
\end{figure}

Shown in Figure~\ref{fig_v2_vs_pt_comparehydro} are the published
experimental data~\cite{Adare:2006ti,Adams:2003am,Adams:2005zg} on
elliptic flow $v_2$ as a function of transverse momentum for various
hadron species in minimum bias $Au+Au$ collisions at $\sqrt{s_{NN}} =
200$~GeV.  Also shown are the results of a particular hydrodynamic
calculation terminated with Cooper-Frye
freeze-out~\cite{Huovinen:2005gy}.  This calculation begins with an
equation of state that assumes chemical equilibrium until kinetic
freeze-out (EoS Q), and a first order phase transition at $T_c$=165
MeV. The initial time is taken as $\tau$=0.6 fm/$c$, and freeze-out
occurs when the temperature drops to $T_{fo}$=130 MeV.  The particle
spectra and elliptic flow ($v_2$) qualitatively agree with the
experimental data for values of $p_T < 1.5$ GeV/$c$.

The reasonable agreement of these calculations with the data, even
without an extensive parameter sensitivity or error analysis, has
generally been taken as evidence that collisions of nuclei at RHIC
form a perfect inviscid fluid (i.e. small $\eta/s$) with short
thermalization times ($\tau_0<0.6$~fm/c).  Notably, different
calculations with different equations of state or initial conditions
do not all agree with the data at the same level~\cite{Adcox:2004mh}.
Similarly small values of $\eta/s$ have also been derived via
measurements of charm suppression and flow~\cite{Adare:2006nq}.  Of
course, the quantitative importance of viscosity has only started to
get serious treatment, both experimentally and
theoretically~\cite{Romatschke:2007mq,Teaney:2003kp,Frodermann:2007ab}.

It is common in the field for the full range of transverse momenta to
be sub-divided into three categories:  
The ``low $p_T$'' region ($p_{T} < 1.5$~GeV) 
where hydrodynamics agrees with
experimental data;  the ``intermediate $p_T$'' region 
($p_{T}\approx1.5-4.5$~GeV/$c$) where hydrodynamics 
over-predicts elliptic flow $v_2$;
and the ``high $p_T$'' region ($p_T>4.5$~GeV) where hadronization from 
jet fragmentation is thought to be dominant.
In the ``intermediate $p_T$'' region, the violation of ``hydrodynamic scaling'' and the observation of
enhanced (anti) baryon relative to meson yields 
(often referred to as the ``baryon
anomaly''~\cite{Fries:2003kq,Adcox:2004mh,Adams:2005dq}) led some to
postulate a different hadronization mechanism.

\section{Recombination and ``$n_q$'' Scaling}

It has recently been speculated that in RHIC
collisions~\cite{Molnar:2003ff} a thermal distribution of quarks might
coalesce (a process sometimes referred to as ``recombination'')
into hadrons when they satisfy the minimum quantum number
requirements.  If these requirements are met in a local region of
phase space, then only the valence quarks of
the correct flavors (and no additional constraints) are needed to form
a hadron.  In~\cite{Fries:2003kq}, they argue
that in the light-cone gauge the hadron wave-function can be expanded
in a Fock space of quarks and anti-quarks
where the first order term is simply the valence quark content
(e.g. $\left|p\right> = \left|uud\right> + \left|uudd\bar{d}\right> +
\ldots$), and that this is applicable when $p_T \gg m$ for the hadron.
This simple picture has been successful in qualitatively
understanding the ``baryon anomaly'' and flow patterns.  
Interestingly, this ``recombination'' picture appears to involve a
decrease in entropy; although, for ``intermediate
$p_T$'' hadrons one can always posit that compensating additional
entropy is generated in the bulk medium at lower $p_T$. In fact, it has
been suggested in~\cite{Biro:2006sv} that the lattice QCD equation of
state permits isentropic hadronization.

\begin{figure}
\includegraphics[width=\columnwidth]{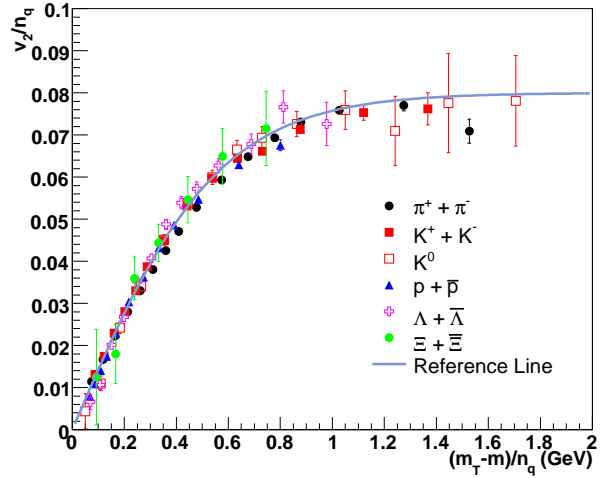}
\caption{\label{fig_v2n_vs_ketn} (color online) The same data as in Figure 1, plotted as 
$v_2$ vs. $m_{T}-m$, the hadron transverse kinetic energy, both scaled by the number
of constituent quarks per hadron.  The reference line is a fit to the kaon data.}
\end{figure}

It has been empirically observed~\cite{Adare:2006ti} that all
hadrons follow a universal trend when $v_2$ versus $p_T$ is
represented instead as $v_2/n_{q}$ versus $(m_{T}-m)/n_{q}$, where
$n_{q}$ is the number of valence quarks in the hadron formed and $m_{T} = \sqrt{p_{T}^{2}+m^{2}}$ 
is the transverse mass.
This was shown with STAR and PHENIX data~\cite{Adare:2006ti} (as reproduced
here) in Figure~\ref{fig_v2n_vs_ketn}, and expressed as a function
of the transverse kinetic energy $KE_{T} = m_{T}-m$.  
This scaling is often termed ``constituent quark scaling.''

If there are quasi-particles with the quantum numbers of quarks that
are coalescing into hadrons, then an immediate question is what are the
values for the thermal masses and widths.
The ``constituent quark'' nomenclature implies an
effective mass of $m_{N}/3 \approx 300$~MeV.
However, the 300 MeV constituent quark mass in a proton is generated by
the spontaneous breaking of chiral symmetry ($\left<q\bar{q}\right>
\ne 0$).  Lattice QCD results indicate an approximate
restoration of chiral symmetry in the quark-gluon plasma region above
the critical temperature~\cite{Allton:2005gk}.  Any such
quasi-particles would be expected to have an effective mass, but not from chiral symmetry breaking.  
Therefore the light quark
masses assigned in~\cite{Greco:2007nu} seem a remarkable coincidence
since the dynamical mass of order $gT$ that might be generated in the
thermal medium does not violate chiral symmetry and is quite different
from a constituent quark mass.  In addition, since the scaling appears
to work for mesons (including the pions, which are anomalously light
Goldstone bosons), a simple coalescence of a 300 MeV quark and a 300
MeV anti-quark easily violates local energy-momentum conservation.

If the quasi-particles containing different flavor quantum numbers
(e.g. light quark flavor versus strange) have different masses, some
scaling violations are expected~\cite{Bellwied:2007tb}.  However, in
the limit that the mass of such quasi-particles is much less than
their momentum, the scaling with $n_q$ gives no information about the
mass differences. In the case of charm flavor quasi-particles, the
current quark mass difference ($\sim 1.4$~GeV) is always large compared with
$gT$ and the momentum up to intermediate $p_T$, so future measurements of
scaling agreement or violation are particularly
interesting~\cite{Bellwied:2007tb,Lin:2003jy}.

\begin{figure}
\includegraphics[width=\columnwidth]{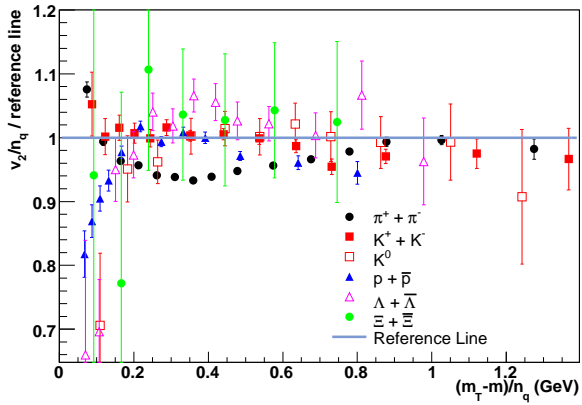}
\caption{\label{fig_v2nreference_vs_ketn} (color online) The same data as in Figure 2, divided
by a reference function fit as described in the text.}
\end{figure}

Despite the categorization of the dominant physics in
different $p_T$ ranges, the scaling does seem to work well at all
$p_T$.  To examine this quantitatively, the data is fit by a reference
curve of the form $v_2/n_q = \alpha \tanh(\beta (m_{T}-m)/n_q)$ (as
shown in Figure~\ref{fig_v2n_vs_ketn}), where the parameters $\alpha$
and $\beta$ are determined from the charged kaon data alone.  Fitting
the kaon data is arbitrary and thus when dividing by the reference
line, one only obtains a measure of the magnitude of deviation of the
different hadrons from each other as shown in
Figure~\ref{fig_v2nreference_vs_ketn}.

The scaling behavior for $(m_{T}-m)/n_{q}$ appears to hold at the $\pm
5$\% level for all hadron species measured for $(m_{T}-m)/n_{q} > 0.2$
GeV.  Note that $(m_{T}-m)/n_{q} = 0.2$ GeV corresponds to $p_T$ =
0.52, 0.75, and 1.21 GeV/$c$ for pions, kaons and protons
respectively.  Below this value, the deviations are larger (e.g. $\pm$
\%15--25\%); It is notable that the charged and neutral kaons have the
largest deviation from each other, which may indicate some
experimental issues.  We note that there are first measurements of
other hadrons (for example
$\phi$~\cite{Afanasiev:2007tv,Abelev:2007rw} and $\Omega$ and light
nuclei $d$ and $^{3}$He~\cite{Afanasiev:2007tv,Liu:2007wu}), but the
current errors only allow one to confirm general agreement with the
scaling behavior.

\begin{figure}
\includegraphics[width=\columnwidth]{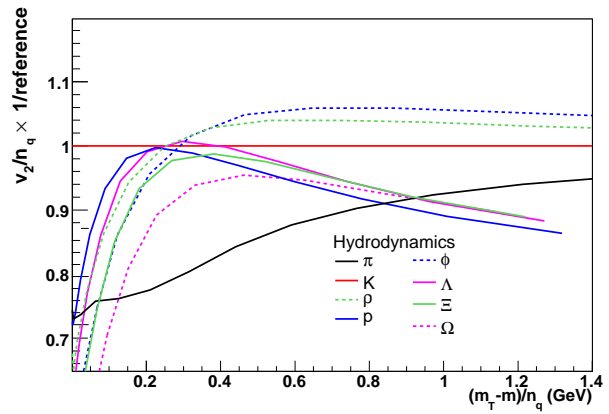}
\caption{\label{hydro2hadronsRef} (color online) The same hydrodynamic calculations shown
in Fig. 1, shown as $v_2/n_q$ vs. $(m_{T}-m)/n_q$.  Note the absence of the
scaling seen in the data.}
\end{figure}

The ``$n_q$'' scaling at low $p_T$ is not a natural prediction
of hydrodynamics followed by Cooper-Frye freeze-out.
These calculations depend only on the particle masses, and
not on the quark content.
However, it has been noted that if
there is a regime where $v_2$ is linear as a function of $m_{T}-m$,
then any re-scaling $v_2/\alpha$ and $(m_{T}-m)/\alpha$ will work for
any constant $\alpha$~\cite{Jia:2006vj}.
In order to test this, in Figure~\ref{hydro2hadronsRef}, a hydrodynamic calculation is
plotted as a function of the scaling variables $v_2/n_q$ and
$(m_T-m)/n_q$.  Again, the different hadron species have been compared
to the kaons, representing an arbitrary choice of reference.
It is notable that the hydrodynamic calculation results are sensitive to the 
choice of initial conditions and equation of state.  However, we do not expect these
variations or the addition of hadronic re-scattering to modify the scaling results.
The results reveal $\pm$15-25\% level deviations of the various 
hadrons from the reference line; whereas, the experimental data shown on the
identical scale in Figure~\ref{fig_v2nreference_vs_ketn} do not.  
The hydrodynamic calculations do not have a linear relation between these
two variables even at low $p_T$, as a function of either $m_{T}-m$ or $p_T$,
at a level that precise experimental data has tested.


As seen in the Figure~\ref{fig_v2nreference_vs_ketn}, it appears that the data follow the
$n_{q}$ scaling hypothesis within experimental uncertainties above $(m_{T}-m)/n_{q} \approx 0.2$ GeV.  
The hydrodynamic calculation shown in Figure~\ref{hydro2hadronsRef} does not, particularly
in the so-called ``hydrodynamic regime.''  This observation does not necessarily imply
that a full recombination calculation would follow the ``$n_q$'' scaling as well as the data, but
it is clear that hydrodynamics fails to predict this emperical observation.


\section{Resonance Contributions}

\begin{figure}
\includegraphics[width=\columnwidth]{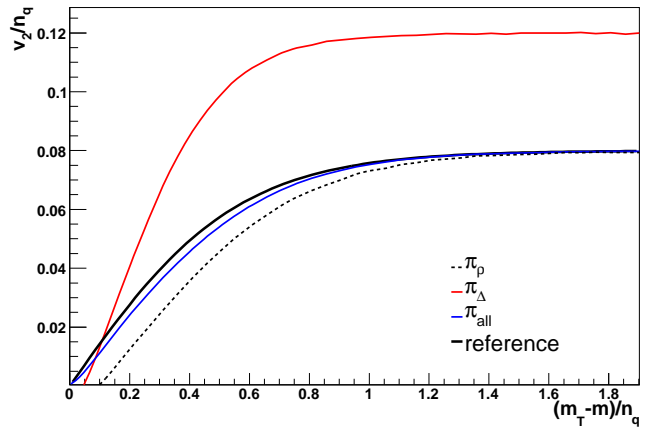}
\caption{\label{rhoDK1} (color online) Simulation results for pion $v_2$ from different sources
are shown.}
\end{figure}

\begin{figure}
\includegraphics[width=\columnwidth]{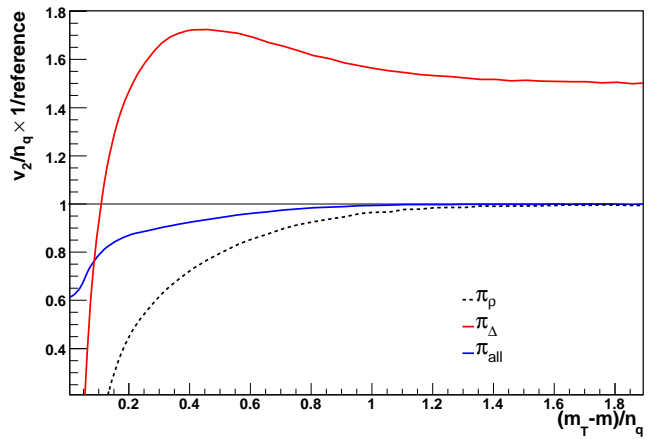}
\caption{\label{rhoDK2} (color online) The results from Figure 5 are divided by the reference line.}
\end{figure}

The level of agreement of the elliptic flow data with ``$n_q$'' scaling is
quite surprising, considering all of the reasons that exist for it not
to work (for example see~\cite{Molnar:2004rr}).  In fact, if all
hadrons are formed via recombination of constituent quarks, then one
might expect that $\rho$ mesons and $\Delta$ resonances would follow
the scaling, but their decay daughter products might not~\cite{Greco:2004ex}.  The decay
of these resonances results in pions with a distorted $v_2$ scaling,
since the decay blurs the emission angle and shifts the $p_T$ of the
pion relative to that of the parent resonance.  We have quantified
this effect using a Monte Carlo simulation. We show in
Figure~\ref{rhoDK1} the $v_2/n_q$ for $\pi$ from $\rho$ decay, $\pi$
from $\Delta$ decay, and pions from all sources combined.  In the
simulation, $\Delta$, $\rho$ and $\pi$ particles are given a $p_T$
according to the spectra calculated in \cite{Huovinen:2005gy} with
hydrodynamics.  In our simulation, the $\rho$ and $\Delta$ resonances,
and the directly hadronized non-decay pions are assumed to follow the
``universal'' scaling curve (reference line) drawn in Figure
\ref{fig_v2n_vs_ketn}.  The $\rho$ and $\Delta$ resonances are
subsequently allowed to decay, and the flow of the pions is computed.
Shown in Figure~\ref{rhoDK2} are the simulation results divided by
the original reference line.  Interestingly, the $\pi$ from $\Delta$
decay has a significantly larger $v_2$ than the scaling, as previously
noted in ~\cite{Greco:2004ex}. In fact, by construction, the elliptic
flow for $\pi$ from $\Delta$ decay must asymptotically approach 1.5
times the thermal pion value at large $p_T$ where the decay products
are essentially co-linear with the parent. This large deviation in the ``$n_q$'' scaling is de-emphasized in the combined result ($\pi_{all}$) due to the small fraction of $\Delta$ decay pions in the sample. 
In contrast, we find that $\pi$ from $\rho$ decay have a slightly
reduced $v_2$ below $(m_{T}-m)/n_q < 0.7$ GeV.  In fact, when the
different contributions are combined, the scaling law appears
``accidentally'' obeyed at the 10\% level with a deviation in a region
of $(m_T -m)/n_q$ similar to that observed in the experimental data. 

\section{Quasi-particle Implications}

Assuming that quasi-particles with the quantum numbers of quarks exist
in the bulk presents some theoretical interpretation challenges since
the light-cone gauge calculation assumes hadrons with $p_T \gg m$ and
moves the entropy issue to lower $p_T$.  In addition, it has been
noted by many authors that the scaling is not an obvious consequence
of hadronization via quark coalescence, but instead a consequence of
the phase space distributions of the recombining quarks, e.g. as shown
in~\cite{Molnar:2004rr}.

One might naively conclude that the medium is a perfect fluid
 and is composed of flowing quasi-particles with
the quantum numbers of quarks (though yet unidentified mass and width
characteristics).  Perhaps these quasi-particles even carry the
thermodynamic degrees of freedom in their particle form.
However, this picture does not sit well with the hydrodynamic paradigm
of early time dynamics at RHIC.  In the case of a perfect fluid, there
can be no well-defined quasi-particles, as their finite mean free path
would induce dissipation and viscous effects.

Alternatively, if the medium is not described by a mathematically {\it perfect} fluid
(i.e. $\eta/s = 0$), but instead
one with a viscosity to entropy density ratio very near the AdS/CFT bound,
$\eta/s\approx 1/4\pi$, one might wonder whether or not it is then possible
for the medium to contain well defined quasi-particles. Although a rigorous answer to this question
is in principle difficult to find, it is possible to derive order
of magnitude estimates for the ratio of a quasi-particle's width
$\Gamma$, to it's mass, $m$ from basic relations. As mentioned previously, a
``well-defined'' quasi-particle is a mode that can be characterized by
a width much smaller than its mass, $\Gamma/m \ll 1$. In the case of a
non-relativistic dilute gas~\cite{Reif:1965}, one finds that the viscosity $\eta$ is
proportional to the product of the average particle momentum, $\langle
p \rangle = m\bar{v}$, the number density, $n$, and the
particle's mean free path, $\lambda$:
\begin{equation}\label{dgaseta}
\eta \sim \frac{1}{3}n\bar{v}m\lambda
\end{equation}
In analogy with the thermodynamics of a non-interacting gas of
massless bosons, the entropy density is assumed to be
proportional to the number density, as $s \sim 4n$.

To introduce the width of the particle, we identify the mean free path
with the particle's lifetime, $\tau$, so that $\lambda=\bar{v}\tau$,
and note that $\Gamma$ and $\tau$ are related by
$\Gamma=2/\tau$. Taking the ratio of $\Gamma$ to $m$ with these
relationships in place suggests that the quasi particle width is
proportional to the temperature, and similar to its mass:
\begin{equation}\label{try1}
\frac{\Gamma}{m}\approx\frac{16T}{3 m}
\end{equation}
where we have used the Maxwell velocity $\bar{v}=\sqrt{8T/\pi m}$ to
characterize the average quasi-particle speed.

An alternative estimate begins from the expression for the viscosity
to entropy density ratio for a weakly coupled quark-gluon plasma
derived from kinetic theory in \cite{Hirano:2005wx},
\begin{equation}\label{ghirano}
\left(\frac{\eta}{s}\right)_{wQGP}=\frac{\lambda T}{5}
\end{equation}
Using again the relationships between $\lambda$, $\tau$, and $\Gamma$,
it happens that the width to mass ratio is once more on the order of the
temperature, this time
\begin{equation}\label{try2}
\frac{\Gamma}{m}\approx \frac{8\pi T}{15m}
\end{equation}
where we have assumed in this case that $\bar{v}\sim c/3 =1/3$. 
In both of these estimates, if we assume that the thermal mass of quasi-particles
is $1-3 \cdot T$, then the ratio $\Gamma/m$ is of order one.  Thus, if the
produced medium has a viscosity near the conjectured bound, 
it is unlikely that quasi-particle modes would be well defined.

%
%
Recently, it has been speculated that the quark-gluon plasma might
have a small viscosity and maintain well defined quasi-particles, if
there is a dominant contribution from turbulent color fields to the
transport coefficients, referred to as an ``anomalous
viscosity''~\cite{Asakawa:2006jn,Asakawa:2006tc}.  This scenario
allows for weakly coupled quasi-particles (and thus well-defined) and
a large collisional viscosity $\eta_{C}$, but where the total
viscosity is determined by the anomalous term, i.e. $\eta^{-1} =
\eta_{A}^{-1} + \eta_{C}^{-1}$.  We note however that the QED plasma
analog is typically non-relativistic, and in the QCD case of the
quark-gluon plasma the system is relativistic. One could imagine that
the heavy flavor elements of such a QGP plasma would be
non-relativistic and thus the compatibility with the QED plasma
calculations would be more obvious.

In a relativistic quantum field theory, the sharp distinction between
fields and particles does not exist and the factorization of viscosity
terms above is non-trivial to mathematically define.  Thus, while it
is very interesting to experimentally determine the possible size of
this anomalous viscosity term, it is unclear if it reconciles the
perfect fluid and quasi-particle pictures.  Applying non-relativistic
calculational insights to the relativistic quark-gluon plasma,
including the previous derivations for $\Gamma/m$, is common.
However, it is often unclear how to determine which physics effects
are actually relevant for describing the observed data.


While quasi-particle widths are large (e.g. while $T>1.05T_c$, which happens 
to be close to the $T_{ch}$ assumed in most calculations), then
hydrodynamics should be the appropriate description of the collision
evolution, as is already assumed for RHIC collisions.  However, as the
temperature approaches $T_c$ from above, the widths may become small
enough at a temperature $T_{qp}$, such that below this, 
a quasi-particle transport (QPT) approach would be applicable, with
quasi-particles following classical paths.
The transition from fluid to quasi-particles would occur at 
the $T=T_{qp}$ hyper-surface with the fluid $v_2$ at that moment.
At the end of this this QPT stage, the quasi-particles
could form hadrons via recombination.


This quasi-particle
transport scenario preceding hadronic transport and then
freeze-out allows the simultaneous use of hydrodynamics (at
$T>T_{qp}$), thermal hadron formation (for $T_{ch}<T<T_{qp}$) and even
hadronic re-scattering (for $T_{th}<T<T_{ch}$, e.g. as implemented in
Ref.\cite{Bass:2000ib}) while allowing in principle for ``$n_q$'' scaling.
With better knowledge of the quasi-particle masses and quantum numbers
it could be possible to implement a multi-stage dynamical model
incorporating all of these with realistic transport.  Of course, the
scaling is not inevitable, but depends quite sensitively on the
details of the dynamical evolution.  For example, if QPT takes place
too early, and freeze-out to hadrons occurs rapidly, then hadronic
re-scattering may well destroy the ``$n_q$'' scaling that could have been
present just after hadron formation.  In addition, if the QPT stage is too short,
the uncertainty principle would in fact preclude narrow width quasi-particles.

\section{Hadronic Transport and ``$n_q$'' Scaling}

In~\cite{Lu:2006qn}, the string and hadron cascade models RQMD and
UrQMD provide another means to understand ``$n_q$'' scaling without
hydrodynamics or degrees of freedom carrying valence quark numbers.
They consider a picture far from the inviscid fluid limit, comprised
of hadrons with long formation times ($\tau \approx 1~fm/c$) and
transport dynamics driven by various {\it in vacuo} interaction cross
sections.  The supposition is that hadron re-scattering encodes the
scaling with valence quark number through different Additive Quark
Model (AQM) hadron-hadron cross sections.  
The use of AQM for hadronic cross sections in these transport models is
directly motivated by the concept that each constituent quark carries a fraction
of the total hadronic cross section.  Thus, the constituent quark number scaling it built
into these cross sections.
It is then immediately seen that ``$n_q$'' scaling does {\it not require} the relevant
degrees of freedom to be carried by separate quark-like objects.
  

Two significant issues pertain to this picture.  First, the overall
magnitude of the $v_2$ is substantially under-predicted by these
calculations.  Of course, if the formation time is reduced, the final
$v_2$ value can be made to approach experimental data.  As
one approaches the hydrodynamic limit where the mean free path is much
shorter than the system size, the relative size of the cross sections
no longer contribute significant differences in flow.  In the
hydrodynamic limit, all cross sections need not be the same, but
rather, the corresponding mean free paths must all be less than
some limiting value.  Additionally, in ~\cite{Lu:2006qn} they excluded the $\phi$ 
and $\Omega$ particles, which have OZI suppressed hadronic cross sections.
In this case, the expected cross section is quite different in the scenario of separate
quark-like objects from that where they are bound together.
Thus, these particles actually represent an excellent test of their hypothesis.
Another additional test would be to show the flow for heavy flavor
particles (such as D mesons)~\cite{Bellwied:2007tb}.

We noted earlier that if one shortens the hadron formation time, one
can reproduce the overall magnitude of elliptic flow. 
If such a calculation were to reproduce the experimental data for all hadrons, it might be tempting to 
conclude the medium dynamics are dominated by hadronic transport, or at least that an ambiguity
exists that precludes ruling out this scenario.
However, in-medium broadening of resonances is likely to occur, which is due to
short time scale interactions (so called ``collision broadening''),
see for example~\cite{Lehr:2001ju}
and references therein. In the case of multiple interactions of resonances on the 
time scale of 0.1 fm/$c$, the widths would be comparable to the masses
and thus the calculation would not be self-consistent in it's treatment
of hadrons in a cascade approach.


\section{Light Nuclei Flow}

\begin{figure}
\includegraphics[width=\columnwidth]{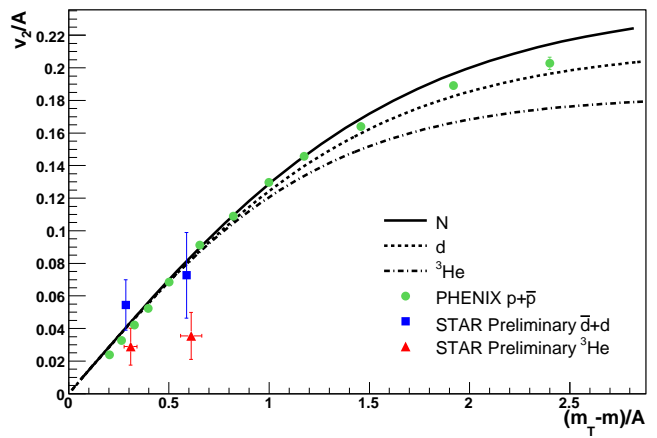}
\caption{\label {nuclei}(color online) Coalescence calculation
comparing $v_2/n_A$ for nucleons (N) and deuterons (d) and $^3$He for minimum bias $Au+Au$ collisions
at $\sqrt{s_{NN}}=200$~GeV. Preliminary
STAR data is shown for d (squares) and $^3$He (triangles) and PHENIX
data is shown for (anti) protons (circles). \cite{Liu:2007wu}.  }
\end{figure}

We have also performed a Monte Carlo coalescence calculation for light
nuclei and their resulting elliptic flow.  In contrast to quark recombination, the mechanism of
nucleosynthesis in heavy ion reactions is well understood and the
light nuclei wave-functions are known.  We parameterize an exploding
fireball freeze-out by a Lorentz boosted distribution of nucleons with
Boltzmann momentum and an induced anisotropic flow. We then coalesce
these nucleons as in ~\cite{Nagle:1996vp}, and compute the flow of
light nuclei shown in Figure~\ref{nuclei}.  The experimental data for
(anti) protons do not follow the reference curve perfectly.  This is because
the reference line is from Figure 2 and is fit to the charged kaons.

The calculation results for deuterons ($^3$He) differ from the simple
scaling of $v_{2}/A$ versus $(m_{T}-m)/A$ of order 10\% (20\%) at high $m_{T}-m$.
In fact, this is expected as can be analytically shown in the case of a delta-function
in momentum space as the coalescence condition.
\begin{equation}
v_2^{d}(p_{T}) = \frac{2v_2^N(p_T/2)}{1+2{v_2^N(p_{T}/2)}^2} \stackrel{v_2^N \ll 1}{\approx} 2v_2^N(p_{T}/2)
\end{equation}
This is the same as the result from analytic quark coalescence
calculations~\cite{Molnar:2003ff,Lin:2003jy}.  Given the large
$v_2$ for protons at large $m_{T}-m$, the correction term in the
denominator exactly accounts for the deviation between the calculated curves shown in Figure~\ref{nuclei}.



In the case of light nuclei coalescence where the wave-function and mechanism are relatively 
well understood, $A$ (or equivalently $n_q$) scaling appears to be reasonably well obeyed simply 
by recombining nucleons close in space-time and momentum.  
In this example it would be incorrect to say that the observed scaling (for example
of light nuclei $v_{2}/n_q$ with $(m_{T}-m)/n_q$ or $v_{2}/A$ with $(m_{T}-m)/A$) stems from baryons 
being the objects flowing with the medium when the flow is built up.
This is not a logical conclusion from the scaling.  Rather, the baryons happen to be flowing with the medium
at a particular time, and the coalescence mechanism imprints that pattern on the light nuclei.

The situation is quite parallel to quark coalescence in that one
cannot conclude that constituent quark degrees of freedom were
relevant as the flow developed.  One only needs these
quasi-particles to have the relevant valence quark degrees of freedom
and to have the right azimuthal anisotropy as the fluid breaks up.  An
interesting question is whether one can at minimum conclude that light
nuclei (in the light nuclei coalescence case) were not the degrees of
freedom at earlier stages. This is not obvious.   Just because the
coalescence mechanism is consistent with the observed phenomena does
not prove that coalescence is the source of light nuclei. For this
conclusion to be valid one would have to show that all the features
present in light nuclei production are consistent with the coalescence
mechanism and that other models for nucleosynthesis fail to describe the
empirical results.
Similarly to the quark coalescence case, can we exclude bound states
(whether hadrons or something more exotic) as the
relevant degrees of freedom at earlier stages?  The previously
discussed UrQMD calculations and uncertainties in the hadronization
mechanism leave this as an open question.


Light nuclei were formed in the early universe by processes such as $p + n \longrightarrow d + \gamma$, where the
$\gamma$ is required to conserve energy and momentum.  In heavy ion reactions these processes have cross sections too small
to account for the yield of light nuclei~\cite{Nagle:1996vp}.  In this case, the coalescence prescription assumes only
that the $p$ and $n$ must overlap with the deuteron wavefunction, and the remaining exchange of energy and/or momentum
with the surrounding medium has little impact.  In fact, deviations from a common penalty factor for each additional coalesced nucleon
has been found to scale with the binding energy per nucleon in the nucleus~\cite{Armstrong:1999xw}.  This may simply reflect
the larger capture cross section for more tightly bound light nuclei, or may be indicative of the required energy and/or momentum
exchange with the medium.  In the case of quark coalescence, the possible required exchange of energy and/or momentum with the 
medium is much larger.  If the quark-like quasi-particles $\tilde{q}$ have for example mass $1.5 T \approx 255~$ MeV, then the formation process
$\tilde{q} + \overline{\tilde{q}} \longrightarrow \pi$ requires a large transfer of energy and/or momentum to the medium.  In fact,
for all hadrons the exchange is much larger than in the case of light nuclei.  At the transition temperature, the mechanism of this
exchange as the chiral condensate is re-appearing is not known.  As pointed out in~\cite{Jia:2006vj}, the scaling of $v_2/n$ with
$m_{T}-m$ as opposed to $p_T$ may be a consequence of conserving energy instead of momentum in this process, though again there is
no {\it a priori} reason for this to be the case.


\section{Summary}

In summary, it has been shown that if the matter produced in heavy ion
collisions is a near-perfect fluid during the stage when the elliptic
flow is built up ($\tau < 5-7$ fm/c), then the active thermodynamic
degrees of freedom during this time period cannot be associated with
specific quasi-particle excitations of the medium.  Thus, the
determination of $\eta/s$ of the medium during this early time stage
is critical to determine whether the medium cannot be
described as composed of quasi-particles.  If there are no
quasi-particles, then the original concept of the quark-gluon plasma
as being composed of quarks and gluons with well-defined thermal
masses is not realized.


By comparing the empirical observation of ``$n_q$'' scaling of
elliptic flow with hydrodynamic fluid calculations followed by
Cooper-Frye hadronization we come to the novel conclusion that at all
$(m_{T}-m)$ ``$n_q$'' scaling gives better agreement to the
experimental data.
It has been pointed out that in the region where $v_2$ is linearly related 
to $(m_{T}-m)$, all values of $\alpha$
would preserve $v_2/\alpha$ as linear with $(m_T-m)/\alpha$.  However,
as shown in Figure ~\ref{hydro2hadronsRef}, typical hydrodynamic
models with Cooper-Frye freeze-out do not evince this linear
relationship.

We also present results for flow scaling for light nuclei and use this
comparison to highlight the similarities and differences of nucleon
coalescence into fragments and quark coalescence into hadrons.
Just as the light nuclei data does not prove that nucleons were the degrees
of freedom when the flow developed, the ``$n_q$'' scaling is not
sufficient evidence that constituent quarks were the degrees of freedom
when the flow developed.

We find that if one had knowledge of the relevant dynamical masses and
quantum numbers of the ensemble of quasi-particles it would be
possible to formulate a multi-stage approach consisting of:
hydrodynamic collective flow, followed by quasi-particle transport
(QPT), coalescence into hadrons, hadronic re-scattering and finally
freeze-out into free streaming hadrons. 
This proposal of a QPT stage between the fluid phase and hadronization
has yet to be theoretically explored.
Most importantly, the masses and widths of
the quasi-particles will have to be related to the creation and
total duration of the QPT phase.

\begin{acknowledgments}
The authors would like to acknowledge our colleagues (B. M\"uller,
D. T. Son, T. Schaefer, and W. A. Zajc) for useful exchanges,
discussions, and suggestions.  The authors also thank P. Huovinen for providing the
hydrodynamic calculation results.  
We acknowledge support from the United States Department of Energy grant
DE-FG02-00ER41152 (L.A.L.L, J.L.N, C.R.) and and DE-AC02-98CH10886 (P.S.).

\end{acknowledgments}
\bibliographystyle{h-physrev3}
\bibliography{cu_npl.bib}


\end{document}